\documentclass[journal]{IEEEtran}
\ifCLASSINFOpdf
\else
\fi
\usepackage{amsmath}
\usepackage{graphicx}
\usepackage{enumitem}
\usepackage{cite}
\usepackage{algorithm}
\usepackage{algorithmic}
\usepackage{multirow}
\usepackage{makecell}
\usepackage{array}
\usepackage{color}
\usepackage{amsfonts}
\newcommand{\tabincell}[2]{\begin{tabular}{@{}#1@{}}#2\end{tabular}}

\begin{document}
%
\title{Triplet-Based Deep Hashing Network for Cross-Modal Retrieval}
%
%
%

\author{Cheng~Deng,
        Zhaojia~Chen,
        Xianglong~Liu,
        Xinbo~Gao,~\IEEEmembership{Senior~Member,~IEEE}
        Dacheng~Tao,~\IEEEmembership{Fellow,~IEEE},
\thanks{C.~Deng, Z.~Chen and X.~Gao are with the School of Electronic Engineering, Xidian University, Xi'an 710071, China (e-mail: chdeng.xd@gmail.com; zhaojiachen@stu.xidian.edu.cn; xbgao@mail.xidian.edu.cn).}
\thanks{X.~Liu is with the State Key Laboratory of Software Development Environment, Beihang University, Beijing 100191, China (e-mail: xlliu@nlsde.buaa.edu.cn).}
\thanks{D.~Tao is with the UBTECH Sydney Artificial Intelligence Centre and the School of Information Technologies, the Faculty of Engineering and Information Technologies, the University of Sydney, 6 Cleveland St, Darlington, NSW 2008, Australia (email: dacheng.tao@sydney.edu.au).}
\thanks{\copyright 20XX IEEE. Personal use of this material is permitted. Permission from IEEE must be obtained for all other uses, in any current or future media, including	reprinting/republishing this material for advertising or promotional purposes, creating new collective works, for resale or redistribution to servers or lists, or reuse of any copyrighted component of this work in other works.}}

%
%

\markboth{}%
{Shell \MakeLowercase{\textit{et al.}}: Bare Demo of IEEEtran.cls for Journals}
%



\maketitle

\begin{abstract}
Given the benefits of its low storage requirements and high retrieval efficiency, hashing has recently received increasing attention. In particular,
cross-modal hashing has been widely and successfully used in multimedia similarity search applications. However, almost all existing methods employing
cross-modal hashing cannot obtain powerful hash codes due to their ignoring the relative similarity between heterogeneous data that contains richer semantic
information, leading to unsatisfactory retrieval performance. In this paper, we propose a triplet-based deep hashing (TDH) network for cross-modal retrieval.
First, we utilize the triplet labels, which describes the relative relationships among three instances as supervision in order to capture more general semantic
correlations between cross-modal instances. We then establish a loss function from the inter-modal view and the intra-modal view to boost the discriminative
abilities of the hash codes. Finally, graph regularization is introduced into our proposed TDH method to preserve the original semantic similarity between
hash codes in Hamming space. Experimental results show that our proposed method outperforms several state-of-the-art approaches on two popular cross-modal datasets.
\end{abstract}


%
\IEEEpeerreviewmaketitle

\section{Introduction}
%
%
%
%
\IEEEPARstart{O}{ver} the last decade, the Internet and social media have developed rapidly, and the volume of multimedia data on the Internet has increased
tremendously. Multimedia data on the Internet exists as a range of different media types and comes from heterogeneous data sources, for example, a webpage may
contain text, audio, images, and video. Although these data are represented by different modalities, they have a strong semantic correlation. Cross-modal retrieval
is designed for scenarios where the queries and retrieval results are from different modalities~\cite{a25,a46}. However, the question of how to effectively bridge
the semantic gap and capture the semantic correlation between heterogeneous data from different modalities is still a challenging one.
\par To reduce the semantic gap, most cross-modal methods, including traditional statistical correlation~\cite{a26,a27,a28}, cross-modal graph
regularization~\cite{a29}\cite{a30}, and dictionary learning~\cite{a31}, are based on subspace learning\cite{a47,a48,a49}, which maps different modality data into
a common subspace and measures the similarities in the common space. However, with the increasing of data, these traditional methods will suffer
from high computing complexity and low search accuracy. To tackle this, hashing-based methods are proposed for large-scale cross-modal retrieval.
Benefiting from low storage cost and high query speed, hashing-based methods have attracted more and more attention from both academia and
industy~\cite{a35,a36,a38}. Cross-modal hashing methods~\cite{a37,a39,a40} transform high-dimensional original data instances into compact binary codes,
with similar binary codes produced for similar data instances, and then compute the Hamming distance among cross-modal data via fast bit-wise XOR operation.
\par Up to now, many hashing-based cross-modal retrieval methods with shallow architectures have been proposed, such as inter-media hashing (IMH)~\cite{a1},
multimodal latent binary embedding (MLBE)~\cite{a32}, collective matrix factorization hashing (CMFH)~\cite{a2}, latent semantic sparse hashing(LSSH)~\cite{a3},
cross-modal similarity-sensitive hashing (CMSSH)~\cite{a4}, and semantic-preserving hashing (SePH)~\cite{a16}. All of these methods are based on hand-crafted
features, which can not effectively capture heterogeneous correlation between different modalities and may therefore result in unsatisfactory performance.
Compared to shallow architectures, deep models can learn multiple levels of representations with increasing abstraction and thus capture heterogeneous cross-modal
correlations more effectively. Deep cross-modal hashing (DCMH)~\cite{a7} simultaneously conducts feature learning and hash code learning in a unified framework.
Pairwise relationship-guided deep hashing (PRDH)~\cite{a11} takes intra-modal and inter-modal constraints into consideration. Deep visual-semantic hashing
(DVSH)~\cite{a33} uses convolutional neural networks (CNNs) and long short-term memory (LSTM) to separately learn unified binary codes for each modality.
However, the text modality in DVSH is constrained to sentences or other sequence texts, which creates limitation in real-world applications.
\par In most of the existing supervised deep cross-modal hashing methods, supervised information is in the form of pairwise labels, which indicates that
two similar instances form a positive pair and two dissimilar instances form a negative one. The loss functions in these methods are designed to preserve the
pairwise similarities of instances. However, the work in~\cite{a7,a11} only preserves the pairwise similarity relations which do not directly encode relative semantic similarity
relationships. Unlike pairwise labels, a triplet-tuple contains a query instance, a positive instance and a negative instance one, where the query
instance is more similar to the positive instance than the negative instance. The instance-similarity relationship is characterized by relative similarity ordering
in the triplets. Furthermore, in contrast to pairwise labels, triplet labels have a critical advantage due to their flexibilities in capturing all kinds
of higher-level similarity, rather than only the binary similar/dissimilar statement as in the case of pairs. Finally, the triplet-like samples can well capture
the intra-class and inter-class variations in the ranking optimization.
\begin{figure*}[!t]
\centering
\includegraphics[width=14cm]{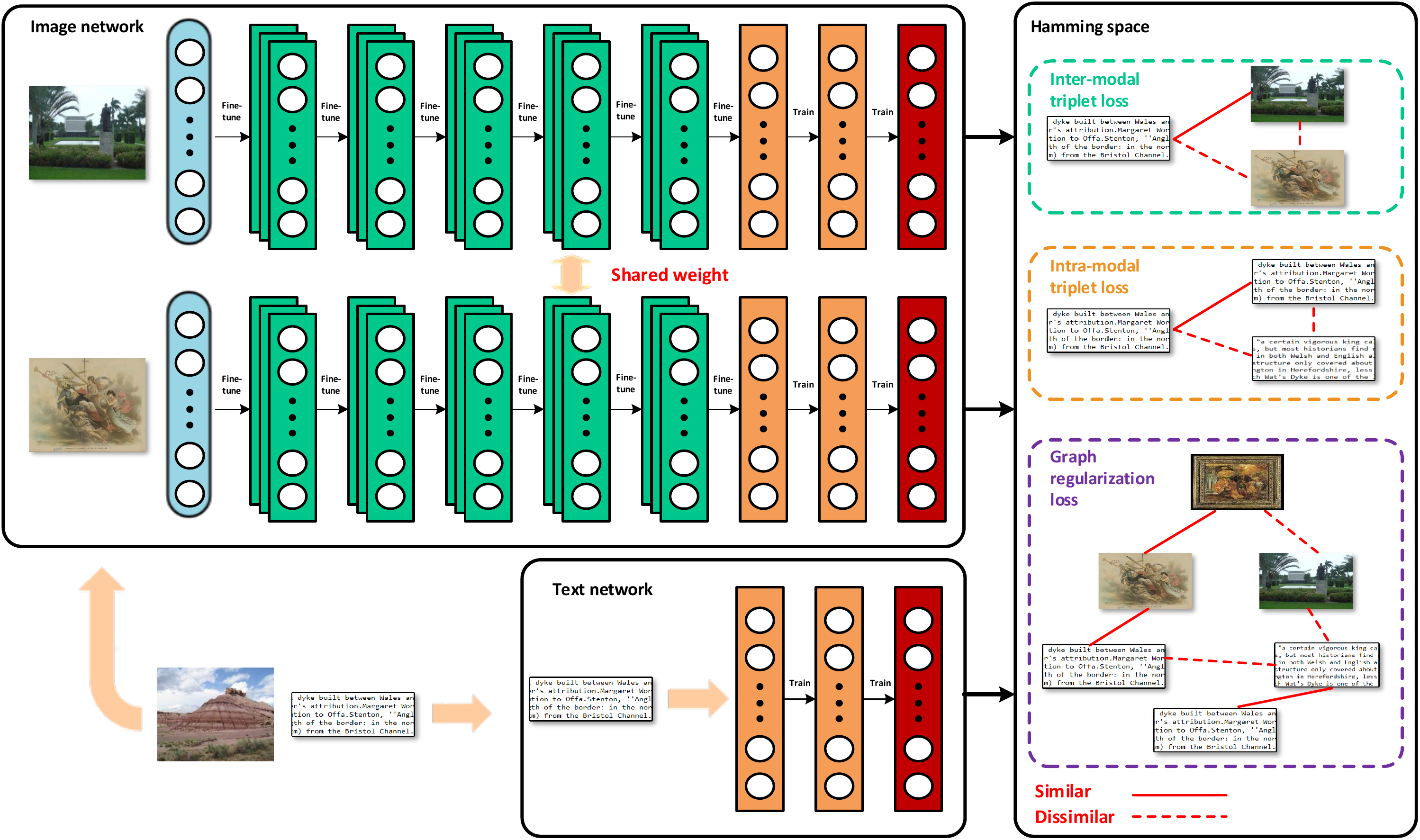}
\caption{The framework of the proposed TDH. We utilize a text as the query instance, while the positive instance and the negative instance are images. The hash codes of the text and the images are obtained by the MLP network and CNN network, respectively.}
\label{fig:graph}
\end{figure*}
\par In this paper, we propose a novel triplet-based deep hashing (TDH) method for large-scale cross-modal retrieval, which uses a deep CNN model to integrate
feature learning and hash code learning for each modality into an end-to-end network. To preserve richer semantic information, we construct triplet labels to
act as supervised information and compute the loss by maximizing the likelihood of the given triplet labels. Although~\cite{a14,a41,a42,a43,a44} also employ
triplet labels, they apply then only to image retrieval whose query data and retrieval results are from the same modalities; Unlike these studies, our proposed
method extends triplet labels to cross-modal retrieval whose query data and retrieval results are from the different modalities. Moreover,
to enhance the discriminative ability of the process, we integrate inter-modality and intra-modality triplet likelihood loss to learn hash codes, such that
the learned hash codes can better reflect the inherent cross-modal correlation. Finally, we introduce graph regularization to preserve the semantic similarities
between hash codes in Hamming space.
Experiments using MIRFlickr25k and NUS-WIDE datasets confirm that our proposed method displays better retrieval performance
than several state-of-the-art approaches.
\par The main contributions of this article can be summarized as follows:
\begin{itemize}
\item We propose a triplet-based deep hashing network (TDH) for cross-modal retrieval, where triplet sampling strategy and specific similarity loss functions are
seamlessly incorporated in an unified framework. The proposed method effectively captures relative semantic similarity relationships and significantly improves retrieval accuracy.

\item The proposed TDH uses triplet labels as supervised information, where triplet labels have a critical advantage due to their flexibilities in capturing all kinds
of higher-level similarity and are prone to generate diverse constraints. Furthermore, triplet organization can enlarge the number of training data to alleviate the over-fitting problem.

\item We utilize triplet labels to construct different triplet loss functions, i.e., the inter-modal triplet loss, the intra-modal triplet loss, and the graph regularization loss, and establish corresponding
similarity relationships of hash codes between original space and Hamming space, which preserves effectively the original semantic similarities between hash codes
in Hamming space and alleviates the semantic gap between cross-modal data.
\end{itemize}
\par The rest of this paper is organized as follows. We introduce some related work pertaining to cross-modal hashing in
Section \uppercase\expandafter{\romannumeral2}. Section \uppercase\expandafter{\romannumeral3} presents our proposed method with theoretical analysis.
Section \uppercase\expandafter{\romannumeral4} describes the algorithm optimization in detail. Section \uppercase\expandafter{\romannumeral5} presents the
experimental results and analysis. Finally, Section \uppercase\expandafter{\romannumeral6} concludes our work.
\section{Related work}
Recently, cross-modal hashing retrieval has attracted considerable attention. According to previous research, cross-modal hashing methods can be roughly
categorized into two groups: unsupervised methods and supervised methods.
\par Unsupervised hashing methods usually learn hash functions from data distribution in order to preserve the structures of training data. For example,
IMH~\cite{a1} explores inter-media consistency and intra-media consistency, and also introduces a linear regression model to jointly learn a set of hash functions
for each distinct modality. CMFH~\cite{a2} utilizes collective matrix factorization to learn latent factor models from different modalities and generate unified
hash codes. Moreover, LSSH~\cite{a3} captures the semantic structures of images via sparse coding and obtains the latent concepts of text via matrix factorization,
then projects the learned semantic feature into a joint common space to obtain unified hash codes.
\par Supervised hashing methods that learn hash functions using supervised information can explore the correlation between heterogeneous data from different
modalities and reduce the semantic gap, which can usually obtain higher accuracy than unsupervised counterparts. CMSSH~\cite{a4} recognizes each hash function as a
binary classification problem and uses a boost algorithm in the learning process. Cross-view similarity search (CVH)~\cite{a5} learns a hash function for each view
when given a set of multi-view training data objects and projects similar objects into similar hash codes across different views. Semantic consistency hashing
(SCM)~\cite{a6} utilizes non-negative matrix factorization and a neighbor preserving algorithm to preserve inter-modal and intra-modal semantic correlations.
\par However, most previous cross-modal hashing methods have been based on shallow architectures, which cannot effectively exploit the complicated heterogeneous
correlation between data across different modalities. In the development of deep neural networks (DNN), deep models are applied to cross-modal
hashing~\cite{a40}\cite{a7,a11,a33} to efficiently capture the correlations between heterogeneous data. Inspired by this idea, we develop a supervised deep model
integrating triplet labels and graph regularization to preserve semantic information and enhance the discriminative ability of the learned hash codes.
\section{Triplet-Based Deep Hashing Network for Cross-Modal Retrieval}
In this section, we introduce our triplet-based deep hashing method (TDH) for cross-modal retrieval in detail, including formulations and learning procedure.
The deep architecture of the TDH model, which integrates feature learning and hashing code learning into a unified end-to-end framework, is shown in
Fig.~\ref{fig:graph}.
\subsection{Notations and Problem Definition}
We first introduce the notations and problem definition for this paper. Our proposed method can be expanded to multiple modalities, such as image, audio and radio.
We use image and text to explain our approach. Bold uppercase letters, such as $\mathbf{X}$, represent matrices; bold lowercase
letters, such as $\mathbf{y}$, are vectors. Moreover, $\mathbf{F}_{*i}$ denotes the $i$th column of $\mathbf{F}$, and $\mathbf{G}^{\top}$ is the transpose
of $\mathbf{G}$. We use $\mathbf{1}$ to denote a vector with all elements being 1. $tr(\cdot)$ and $\lVert\cdot\rVert_F$ represent the trace and the Frobenius
norm of a matrix, respectively. $sign(\cdot)$ is the sign function, as follows:\\
\begin{center}
$sign(x)=\begin{cases}
          \;\,\,\,1\quad & x\ge0\\
         -1\quad & x<0.
         \end{cases}$
\end{center}
\par Assuming that there are $N$ training instances $\mathcal{O} = \{\mathbf{o}_i\}_{i=1}^N$, each of instance has features from two modalities, such as text
and image. We use $\mathbf{X}=\{\mathbf{x}_i\}_{i=1}^N$ and $\mathbf{Y}=\{\mathbf{y}_i\}_{i=1}^N$ to denote textual modality and image modality, respectively.
The $M$ triplet labels are represented as $ \mathcal{T}=\{ (q_1,p_1,n_1),\dotsi,(q_M,p_M,n_M)\}$, where the triplet indice $(q_m,p_m,n_m) $ denotes that the query
instance of index $q_m $ is more similar to the instance of index $p_m $ than to the instance of index $n_m $. We obtain triplet labels by selecting the query
instances $\mathbf{o}_{q_m}$ and the positive instances $\mathbf{o}_{p_m}$, which have the same semantic label, and while the negative instances $\mathbf{o}_{n_m} $ from a different semantic label.
\par Given the training data $\mathbf{X}$, $\mathbf{Y}$ and the triplet labels $\mathcal{T} $, our goal in cross-modal hashing is to learn two hash functions,
$h^{x}(\mathbf{x})\in\{-1,1\}^{k} $ for text and $h^{y}(\mathbf{y})\in\{-1,1\}^{k} $ for images, where $k $ is the length of the hash code. These hash functions
should satisfy the semantic similarity in $\mathcal{T} $. More specifically, $dist_{H}(\mathbf{b}_{q_m}^{x},\mathbf{b}_{p_m}^{y}) $ should be smaller than
$dist_{H}(\mathbf{b}_{q_m}^{x},\mathbf{b}_{n_m}^{y}) $ where $dist_{H}(\cdot,\cdot) $ represents the Hamming distance between two hash codes,
and $\mathbf{b}_{i}^{x}=h^{x}(\mathbf{x}_{i}) $, $\mathbf{b}_{i}^{y}=h^{y}(\mathbf{y}_{i}) $.
\begin{table}
\caption{The CNN architecture for image modality. }
\label{tab:CNN}
\center
\tabcolsep8pt
\arrayrulewidth1pt
\begin{tabular}{|c|c|}
\hline
Layer & Configuration\\
\hline
\hline
conv1 & kernel:64$\times$11$\times$11, stride:4, pad:0, LRN, $\times$2 pool\\
conv2 & kernel:256$\times$5$\times$5, stride:1, pad:2, LRN, $\times$2 pool\\
conv3 & kernel:256$\times$3$\times$3, stride:1, pad:1\\
conv4 & kernel:256$\times$3$\times$3, stride:1, pad:1\\
conv5 & kernel:256$\times$3$\times$3, stride:1, pad:1, $\times$2 pool\\
fc6 & 4096 dropout\\
fc7 & 4096 dropout\\
fch8 & hash code length $k$\\
\hline
\end{tabular}
\end{table}
\begin{table}
\caption{The MLP architecture for textual modality. }
\label{tab:MLP}
\center
\tabcolsep8pt
\arrayrulewidth1pt
\begin{tabular}{|c|c|}
\hline
Layer & Configuration\\
\hline
\hline
fc1 & length of BOW vector\\
fc2 & 4096 dropout\\
fch3 & hash code length $k$\\
\hline
\end{tabular}
\end{table}
\subsection{Deep Architecture}
We apply two deep neural networks, one for image modality and the other for textual modality.
\par Table~\ref{tab:CNN} shows the deep architecture for the image modality. We adopt the CNN-F~\cite{a10} network architecture pre-trained on the ImageNet
dataset~\cite{a22} for image feature learning, due to its excellent performance on object classification. The original CNN-F model contains five convolutional
layers ($conv$) and three fully-connected layers ($fc$). To map the learned deep features into Hamming space directly, we replace the $fc8$ layer with a
fully-connected hash $(fch)$ layer with $k$ hidden nodes, as in~\cite{a11}. The details of each of the layers are given in Table~\ref{tab:CNN}. ``kernel" indicates
the number of convolution filters and their receptive field size, while ``stride" and ``pad" are the convolution stride and padding respectively. ``LRN" denotes
whether Local Response Normalization~\cite{a34} is applied or not. ``pool" is the down-sampling factor.
\par For the textual modality, we first transform each text into a vector using bag-of-words (BOW) representation. The BOW is used as the input to the Multi-Layers
Perceptron (MLP) to extract deep textual features. The MLP consists of three fully-connected layers whose last layer is replaced with a new $fch$
hash layer with $k$ hidden nodes, in the same way as for the image modality in~\cite{a11}. The details of each of the $fc$ layers are provided in
Table~\ref{tab:MLP}. We can use other deep networks for feature learning, such as AlexNet~\cite{a12} and GoogleNet~\cite{a13}, which will be studied in the future.
\subsection{The Triplet Loss}
 The triplet loss boosts the correlation among three instances $\{\mathbf{o}_{q_m},\mathbf{o}_{p_m},\mathbf{o}_{n_m}\}$ and reduces the similarity between
 semantically dissimilar instances $\{\mathbf{o}_{q_m},\mathbf{o}_{n_m}\}$ while increasing the similarity between semantically similar instances
 $\{\mathbf{o}_{q_m},\mathbf{o}_{p_m}\}$. In more detail, we utilize triplet label negative log likelihoods~\cite{a14} to model such relationships among three
 instances.
\par Suppose that the positive instance and the negative instance are text, and the query instance is an image. $\mathbf{F}\in\mathbb{R}^{k\times N} $ denotes
the hash codes of image modality output from the image neural network, $\mathbf{G}\in\mathbb{R}^{k\times N}$ denotes the hash codes of textual modality output
from the textual neural network, and $\mathcal{T} $ represents the triplet labels. Accordingly, the triplet label likelihood is defined as:
\begin{equation}
\label{form:triplet loss}
p(\mathcal{T}|\mathbf{F,G,G}) = \prod\nolimits^{M} \limits_{m=1}p((q_m,p_m,n_m)|\mathbf{F,G,G}).
\end{equation}
with
\begin{equation}
p((q_m,p_m,n_m)|\mathbf{F,G,G})= \sigma(\theta_{q_m^yp_m^x}-\theta_{q_m^yn_m^x}-\alpha),
\end{equation}
where $\theta_{q_m^yp_m^x}=\frac{1}{2}\mathbf{F}^{\top}_{*q_m}\mathbf{G}_{*p_m}$, $\theta_{q_m^yn_m^x}=\frac{1}{2}\mathbf{F}^{\top}_{*q_m}\mathbf{G}_{*n_m}$,
$\sigma(x) $ is the sigmoid function $\sigma(x) = \frac{1}{1+e^{-x}}$, and the threshold $\alpha$ is a margin that is enforced between positive and negative pairs,
a hyper-parameter. $\mathbf{F}^{\top}_{*i} = f^y(\mathbf{y}_i;w_y) $, and $\mathbf{G}^{\top}_{*i}=f^x(\mathbf{x}_i;w_x) $, where $w_x$, $w_y$ are the network
parameters of textual modality and image modality, respectively.
\par To better preserve the semantic similarities of training instances in Hamming space and enhance discriminability of the learned hash codes, we construct
the objective function with three parts: (1) Inter-modal triplet loss; (2) Intra-modal triplet loss; (3) Graph regularization loss.
\subsection{Inter-Modal Triplet Loss}
\par For efficient cross-modal hashing retrieval, we add the inter-modal triplet embedding loss to effectively capture the heterogeneous correlations across
different modalities. Therefore, the inter-modal triplet loss for Image-to-Text is formulated as:
\begin{equation}
\begin{aligned}
J_1 & \!=\!-log\ p(\mathcal{T}|\mathbf{F,G,G})\\
& \!=\!-\!\sum\nolimits^{M}\limits_{m=1}\!log\ p((q_m,p_m,n_m)|\mathbf{F,G,G})\\
& \!=\!-\!\sum\nolimits^{M}\limits_{m=1}\!(\theta_{q_m^yp_m^x}\!-\!\theta_{q_m^yn_m^x}\!-\!\alpha\!-log(1\!+\!e^{\theta_{q_m^yp_m^x}\!-\!\theta_{q_m^yn_m^x}\!-\!\alpha})) ,
\end{aligned}
\end{equation}
The inter-modal triplet loss for Text-to-Image is formulated as:
\begin{equation}
\begin{aligned}
J_2 & \!=\!-log\ p(\mathcal{T}|\mathbf{G,F,F})\\
& \!=\!-\!\sum\nolimits^{M}\limits_{m=1}\!log\ p((q_m,p_m,n_m)|\mathbf{G,F,F})\\
& \!=\!-\!\sum\nolimits^{M}\limits_{m=1}\!(\theta_{q_m^xp_m^y}\!-\!\theta_{q_m^xn_m^y}\!-\!\alpha\!-\!log(1\!+\!e^{\theta_{q_m^xp_m^y}\!-\!\theta_{q_m^xn_m^y}\!-\!\alpha})) ,
\end{aligned}
\end{equation}
where $\theta_{q_m^xp_m^y}=\frac{1}{2}\mathbf{G}^{\top}_{*q_m}\mathbf{F}_{*p_m} $ and $\theta_{q_m^xn_m^y}=\frac{1}{2}\mathbf{G}^{\top}_{*q_m}\mathbf{F}_{*n_m} $. Thus the inter-modal triplet loss is defined as:
\begin{equation}
\label{inter}
J_{inter} = J_1 + J_2 .
\end{equation}
\par It is evident that optimizing the above loss will reduce the Hamming distance between the query $\mathbf{F}_{q_m} $/$\mathbf{G}_{q_m} $ and the
positive instances $\mathbf{G}_{p_m}$/$\mathbf{F}_{p_m} $ while increasing the Hamming distance between the query $\mathbf{F}_{q_m} $/$\mathbf{G}_{q_m} $ and
the negative instances $\mathbf{G}_{n_m} $/$\mathbf{F}_{n_m} $. This can preserve the semantic correlation among instances from different modalities.
\subsection{Intra-Modal Triplet Loss}
\par To obtain more accurate hash codes, we focus on not only preserving cross-modal semantic similarity, but also on exploring the discriminative abilities
of hash codes in their own modality to preserve the semantic information, which can improve the performance of cross-modal retrieval. Accordingly, it is necessary
to add the intra-modal triplet loss into the objective function. For image modality, the intra-modal triplet loss is formulated as follows:
\begin{equation}
\begin{aligned}
J_3 & \!=\!-log\ p(\mathcal{T}|\mathbf{F})\\
& \!=\!-\!\sum\nolimits^{M}\limits_{m=1}\!log\ p((q_m,p_m,n_m)|\mathbf{F})\\
& \!=\!-\!\sum\nolimits^{M}\limits_{m=1}\!(\theta_{q_m^yp_m^y}\!-\!\theta_{q_m^yn_m^y}\!-\!\alpha\!-\!log(1\!+\!e^{\theta_{q_m^yp_m^y}\!-\!\theta_{q_m^yn_m^y}\!-\!\alpha})) ,
\end{aligned}
\end{equation}
where $\theta_{q_m^yp_m^y}=\frac{1}{2}\mathbf{F}^{\top}_{*q_m}\mathbf{F}_{*p_m} $ and $\theta_{q_m^yn_m^y}=\frac{1}{2}\mathbf{F}^{\top}_{*q_m}\mathbf{F}_{*n_m} $.\\
For textual modality, the intra-modal triplet loss is formulated homoplastically as:
\begin{equation}
\begin{aligned}
& J_4\!=\!-log\ p(\mathcal{T}|\mathbf{G})\\
& \!=\!-\!\sum\nolimits^{M}\limits_{m=1}\!log\ p((q_m,p_m,n_m)|\mathbf{G})\\
& \!=\!-\!\sum\nolimits^{M}\limits_{m=1}\!(\theta_{q_m^xp_m^x}\!-\!\theta_{q_m^xn_m^x}\!-\!\alpha\!-\!log(1\!+\!e^{\theta_{q_m^xp_m^x}\!-\!\theta_{q_m^xn_m^x}\!-\!\alpha})) ,
\end{aligned}
\end{equation}
where $\theta_{q_m^xp_m^x}=\frac{1}{2}\mathbf{G}^{\top}_{*q_m}\mathbf{G}_{*p_m} $ and $\theta_{q_m^xn_m^x}=\frac{1}{2}\mathbf{G}^{\top}_{*q_m}\mathbf{G}_{*n_m} $. Thus, the intra-modal triplet loss is:
\begin{equation}
\label{intra}
J_{intra} = J_3 + J_4 .
\end{equation}
\subsection{Graph Regularization Loss}
\par The nearest neighbor graph is able to reflect the degree of similarity among multiple data points. We introduce graph regularization into the Hamming space
to enhance the correlation between the unified hash codes. We formulate a spectral graph learning problem from the label similarity matrix $\mathbf{S}$ as:
\begin{center}
$\frac{1}{2}\sum\nolimits^{N}\limits_{i,j=1}\lVert\mathbf{b}_i-\mathbf{b}_j\rVert^2\mathbf{S}_{ij}=tr(\mathbf{B L}\mathbf{B}^{\top}),$
\end{center}
where $\mathbf{S} $ is the similarity matrix and $\mathbf{B}=\{\mathbf{b}_i\}_{i=1}^N $ represents the unified hash codes. If $\mathbf{o}_i $ and $\mathbf{o}_j $
have the same labels, $s_{ij}=1 $; otherwise, $s_{ij}=0 $. We define $\mathbf{D} = diag(d_1,\dots,d_n) $, and $\mathbf{L=D-S} $ is the Laplacian matrix. Therefore
the graph regularization loss can be formulated as:
\begin{equation}
\begin{aligned}
\label{re}
J_{re} = & \gamma (\lVert \mathbf{B}^{y}-\mathbf{F}\rVert_F^2+\lVert \mathbf{B}^{x}-\mathbf{G}\rVert_F^2)\\
& +\eta(\lVert \mathbf{F}\cdot\mathbf{1}\rVert_F^2+\lVert \mathbf{G}\cdot\mathbf{1}\rVert_F^2)+\beta tr(\mathbf{BLB}^{\top})\\
& s.t.\quad \mathbf{B}=\mathbf{B}^{x}=\mathbf{B}^{y}\in \{-1,1\}^{k\times N},
\end{aligned}
\end{equation}
where $\mathbf{B}^{x}$ is the hash codes of textual modality and $\mathbf{B}^{y}$ is the hash codes of image modality. The first and second terms represent the
quantization error to solve the relaxed problem. Simultaneously, the third and fourth terms are used to make the balanced bit such that the number of 1 and $-1$
for each bit on the hash codes should be nearly the same. $\gamma$, $\eta $ and $\beta$ are parameters employed to balance the weight of each part.
\subsection{Objective Function}
\par By merging the above three parts together (i.e., the inter-modal triplet loss $J_{inter}$, the intra-modal triplet loss $J_{intra}$, and the graph
regularization loss $J_{re}$), we obtain the whole objective function, formulated as the following minimization problem:
\begin{equation}
\label{form:obj}
\min\limits_{\mathbf{B},w_x,w_y}J=\min\limits_{\mathbf{B},w_x,w_y}J_{inter}+J_{intra}+J_{re}.
\end{equation}
\section{Learning Algorithm}
In this section, we introduce the optimization algorithm in detail. Since the objective function is not convex, we adopt the mini-batch Stochastic Gradient
Descent (SGD) method and alternating learning strategy to learn $w_x$, $w_y$ and $\mathbf{B}$.
We update one parameter with each time, while other parameters remain fixed. The model is optimized
iteratively until parameters converge or the preset maximum number of iterations is reached. The optimization procedure of our cross-modal hashing learning
model is summarized in Algorithm~\ref{alg:A}.
\subsection{Updating $\mathbf{B}$}
When $w_x$ and $w_y$ are fixed, the objective function in~\eqref{form:obj} can be expanded as follows:
\begin{equation}
\begin{aligned}
\label{form:B}
& \min\limits_{\mathbf{B}}\gamma tr(\mathbf{B}^{\top}{\mathbf{B}}-\mathbf{F}\mathbf{B}^{\top}-\mathbf{G}\mathbf{B}^{\top})+\beta tr(\mathbf{BLB}^{\top})\\
& s.t.\quad\mathbf{B}\in \{ -1,1\}^{k\times N}.
\end{aligned}
\end{equation}
We compute the derivation of~\eqref{form:B} with respect to $\mathbf{B}$ and infer that $\mathbf{B}$ should be defined as follows:
\begin{equation}
\label{form:update B}
\mathbf{B}=sign((\mathbf{F+G})(2\mathbf{I}+\frac{\beta}{\gamma}\mathbf{L})^{-1}),
\end{equation}
where $\mathbf{I}$ denotes the identity matrix.
\begin{algorithm}
\caption{Optimization procedure for TDH}
\label{alg:A}
\begin{algorithmic}
\REQUIRE
\STATE{Text set $\mathbf{X}$, image set $\mathbf{Y}$, and the set of triplet labels $\mathcal{T}$.}
\ENSURE
\STATE{Parameters $w_x$ and $w_y$ of the deep neural networks, and binary code matrix $\mathbf{B}$.\\
\textbf{Initialization}\\
Initialize neural parameters $w_X$ and $w_y$, mini-batch size $N_x=N_y=128$, and iteration number $t_x=N/N_x$, $t_y=N/N_y$.}
\REPEAT
\STATE{Update $\mathbf{B}$ according to~\eqref{form:update B}.}
\FOR{iter=$1,2,\cdots,t_x$}
\STATE{Randomly sample $N_x$ instances from $\mathbf{X}$ to construct a mini-batch $\mathbf{X}_{N_x}$ and make up a triplet set where the query instances
come from $\mathbf{X}_{N_x}$.\\
For each sampled instances $\mathbf{x}_i$ in the mini-batch, calculate $\mathbf{G}_{*i}=f(\mathbf{x}_i;w_x)$ by forward propagation.\\
Calculate the derivative according to~\eqref{form:update G}.\\
Update parameter $w_x$ using back propagation.}
\ENDFOR
\FOR{iter=$1,2,\cdots,t_y$}
\STATE{Randomly sample $N_y$ instances from $\mathbf{Y}$ to construct a mini-batch $\mathbf{Y}_{N_y}$ and make up a triplet set where the query instances
come from $\mathbf{Y}_{N_y}$.\\
For each sampled instances $\mathbf{y}_i$ in the mini-batch, calculate $\mathbf{F}_{*i}=f(\mathbf{y}_i;w_y)$ by forward propagation.\\
Calculate the derivative according to~\eqref{form:update F}.\\
Update parameter $w_y$ using back propagation.}
\ENDFOR
\UNTIL{a fixed number of iterations;}
\end{algorithmic}
\end{algorithm}
\subsection{Updating $w_x$}
When $w_y$ and $\mathbf{B}$ are fixed, the neural network parameter $w_x$ is learned using SGD with a Back Propagation (BP) algorithm. In order to comprehend
the optimization, we divide the gradient of the loss into inter-modal gradient, intra-modal gradient and regularization gradient, respectively. For the $i$th
instance $\mathbf{G}_{*i}$, we derive the gradient of the loss as follows:
\begin{small}
\begin{equation}
\begin{aligned}
\label{form:update G}
\frac{\partial J}{\partial\mathbf{G}_{*i}}= & \frac{\partial J_{inter}}{\partial\mathbf{G}_{*i}}+\frac{\partial J_{intra}}{\partial\mathbf{G}_{*i}}+\frac{\partial J_{re}}{\partial\mathbf{G}_{*i}}\\
= & -\!\frac{1}{2}\!\sum\nolimits^{M}\limits_{m:(i,p_m,n_m)}\!(1\!-\!\sigma(\theta_{ip_m^y}\!-\!\theta_{in_m^y}\!-\!\alpha))(\mathbf{F}_{*p_m}\!-\!\mathbf{F}_{*n_m})\\
& -\!\frac{1}{2}\!\sum\nolimits^{M}\limits_{m:(i,p_m,n_m)}\!(1\!-\!\sigma(\theta_{ip_m^x}\!-\!\theta_{in_m^x}\!-\!\alpha))(\mathbf{G}_{*p_m}\!-\!\mathbf{G}_{*n_m}\!)\\
& +2\gamma(\mathbf{G-B})+2\eta\mathbf{G1}.
\end{aligned}
\end{equation}
\end{small}
\subsection{Updating $w_y$}
We also train the neural network parameter $w_y$ using the SGD algorithm and divide the gradient into three parts. For the $i$th instance
$\mathbf{F}_{*i}$, we compute the derivative of the loss function as:
\begin{small}
\begin{equation}
\begin{aligned}
\label{form:update F}
\frac{\partial J}{\partial\mathbf{F}_{*i}}= & \frac{\partial J_{inter}}{\partial\mathbf{F}_{*i}}+\frac{\partial J_{intra}}{\partial\mathbf{F}_{*i}}+\frac{\partial J_{re}}{\partial\mathbf{F}_{*i}}\\
= & -\!\frac{1}{2}\!\sum\nolimits^{M}\limits_{m:(i,p_m,n_m)}\!(1\!-\!\sigma(\theta_{ip_m^x}\!-\!\theta_{in_m^x}\!-\!\alpha))(\mathbf{G}_{*p_m}\!-\!\mathbf{G}_{*n_m})\\
& -\!\frac{1}{2}\!\sum\nolimits^{M}\limits_{m;(i,p_m,n_m)}\!(1\!-\!\sigma(\theta_{ip_m^y}\!-\!\theta_{in_m^y}\!-\!\alpha))(\mathbf{F}_{*p_m}\!-\!\mathbf{F}_{*n_m}\!)\\
& +2\gamma(\mathbf{F-B})+2\eta\mathbf{F1}.
\end{aligned}
\end{equation}
\end{small}
\par In each iteration, we sample a mini-batch of data from the training set and implement our alternating learning strategy on the sample data. More specifically,
if we sample $P$ data each time, we are able to obtain $3\times P$ data from the triplet labels which is equivalent to a twofold increase in the amount of training
data and therefore improves the training performance.
\subsection{Out-of-Sample Extension}
Any new instance that is not from the training data can be represented as a specific hash code as long as one of its modalities (text or image) are observed. For
example, given one instance $q$ from textual modality $\mathbf{x}_q$, we can generate a hash code by forward propagating the MLP network, as follows:
\begin{center}
$\mathbf{b}_q^x=h^x(\mathbf{x}_q)=sign(f(\mathbf{x}_q;w_x)).$
\end{center}
Similarly, given an instance $q$ from image modality $\mathbf{y}_q$, we can also generate the hash code by CNN-F network, as follows:
\begin{center}
$\mathbf{b}_q^y=h^y(\mathbf{y}_q)=sign(f(\mathbf{y}_q;w_y)).$
\end{center}
\par In this way, it can be seen that our proposed method can be utilized for cross-modal retrieval where the query data and the result data are from different
modalities.
\begin{table}
\caption{Comparison to baselines with hand-crafted features on MIRFLICKR-25K in terms of MAP. Best accuracy is shown in boldface.}
\label{tab:flickr_hand}
\center
\tabcolsep8pt
\arrayrulewidth1pt
\begin{tabular}{|c|c|c|c|c|}
\hline
\multirow{2}{*}{Task/MIRFlickr25k}&
\multirow{2}{*}{Methods}&
\multicolumn{3}{c|}{Code Length}\\
\cline{3-5}
&&16 bits&32 bits&64 bits\\
\hline
\hline
\multirow{10}{*}{\tabincell{c}{Image Query\\v.s.\\Text Database}}&CMFH~\cite{a2}&0.5804&0.5790&0.5797\\
\cline{2-5}
&SCM~\cite{a6}&0.6153&0.6279&0.6288\\
\cline{2-5}
&LSSH~\cite{a3}&0.5784&0.5804&0.5797\\
\cline{2-5}
&STMH~\cite{a20}&0.5876&0.5951&0.5942\\
\cline{2-5}
&CVH~\cite{a5}&0.6067&0.6177&0.6157\\
\cline{2-5}
&SePH~\cite{a16}&0.6441&0.6492&0.6508\\
\cline{2-5}
&DCMH~\cite{a7}&0.7056&0.7035&0.7140\\
\cline{2-5}
&PRDH~\cite{a11}&0.6819&0.6917&0.6913\\
\cline{2-5}
&\textbf{TDH}&\textbf{0.7110}&\textbf{0.7228}&\textbf{0.7289}\\
\hline
\hline
\multirow{10}{*}{\tabincell{c}{Text Query\\v.s.\\Image Database}}&CMFH~\cite{a2}&0.5728&0.5778&0.5779\\
\cline{2-5}
&SCM~\cite{a6}&0.6102&0.6184&0.6192\\
\cline{2-5}
&LSSH~\cite{a3}&0.5898&0.5927&0.5932\\
\cline{2-5}
&STMH~\cite{a20}&0.5763&0.5877&0.5826\\
\cline{2-5}
&CVH~\cite{a5}&0.6026&0.6041&0.6017\\
\cline{2-5}
&SePH~\cite{a16}&0.6455&0.6474&0.6506\\
\cline{2-5}
&DCMH~\cite{a7}&0.7311&0.7487&0.7499\\
\cline{2-5}
&PRDH~\cite{a11}&0.7340&0.7397&0.7418\\
\cline{2-5}
&\textbf{TDH}&\textbf{0.7422}&\textbf{0.7500}&\textbf{0.7548}\\
\hline
\end{tabular}
\end{table}
\begin{table}
\caption{Comparison to baselines with hand-crafted features on NUS-WIDE in terms of MAP. Best accuracy is shown in boldface.}
\label{tab:NUS-WIDE_hand}
\center
\tabcolsep8pt
\arrayrulewidth1pt
\begin{tabular}{|c|c|c|c|c|}
\hline
\multirow{2}{*}{Task/NUS-WIDE}&
\multirow{2}{*}{Methods}&
\multicolumn{3}{c|}{Code Length}\\
\cline{3-5}
&&16bits&32bits&64bits\\
\hline
\hline
\multirow{10}{*}{\tabincell{c}{Image Query\\v.s.\\Text Database}}&CMFH~\cite{a2}&0.3825&0.3858&0.3890\\
\cline{2-5}
&SCM~\cite{a6}&0.4904&0.4945&0.4992\\
\cline{2-5}
&LSSH~\cite{a3}&0.3900&0.3924&0.3962\\
\cline{2-5}
&STMH~\cite{a20}&0.4344&0.4461&0.4534\\
\cline{2-5}
&CVH~\cite{a5}&0.3687&0.4182&0.4602\\
\cline{2-5}
&SePH~\cite{a16}&0.5314&0.5340&0.5429\\
\cline{2-5}
&DCMH~\cite{a7}&0.6141&0.6167&0.6427\\
\cline{2-5}
&PRDH~\cite{a11}&0.5874&0.6154&0.6232\\
\cline{2-5}
&\textbf{TDH}&\textbf{0.6393}&\textbf{0.6626}&\textbf{0.6754}\\
\hline
\hline
\multirow{10}{*}{\tabincell{c}{Text Query\\v.s.\\Image Database}}&CMFH~\cite{a2}&0.3915&0.3944&0.3990\\
\cline{2-5}
&SCM~\cite{a6}&0.4595&0.4650&0.4691\\
\cline{2-5}
&LSSH~\cite{a3}&0.4286&0.4248&0.4248\\
\cline{2-5}
&STMH~\cite{a20}&0.3845&0.4089&0.4181\\
\cline{2-5}
&CVH~\cite{a5}&0.3646&0.4024&0.4339\\
\cline{2-5}
&SePH~\cite{a16}&0.5086&0.5055&0.5710\\
\cline{2-5}
&DCMH~\cite{a7}&0.6591&0.6487&\textbf{0.6847}\\
\cline{2-5}
&PRDH~\cite{a11}&0.6303&0.6432&0.6568\\
\cline{2-5}
&\textbf{TDH}&\textbf{0.6647}&\textbf{0.6758}&0.6803\\
\hline
\end{tabular}
\end{table}
\subsection{Multiple Modalities Extension}
The proposed method can be extended to more modalities, and the network structures are extended to more networks.
The extension of TDH in~\eqref{form:obj} from bimodal to multiple modalities is quite simple and direct as follow:
\begin{small}
\begin{equation}
\begin{aligned}
\min\limits_{\mathbf{B},w_1,\dots,w_R}&\sum\nolimits^{R}\limits_{i,j=1,i\not= j}J_{inter}^{ij}+\sum\nolimits^{R}\limits_{i=1}J_{intra}^i\\
&+\sum\nolimits^{R}\limits_{i=1}(\gamma\lVert\mathbf{B}^i-\mathbf{H}^i\rVert_F^2+\eta\lVert\mathbf{H}^i\cdot\mathbf{1}\rVert_F^2)+\beta tr(\mathbf{BLB}^{\top})\\
&s.t.\quad \mathbf{B}=\mathbf{B}^1=\dots=\mathbf{B}^R\in\{-1,1\}^{k\times N}.
\end{aligned}
\end{equation}
\end{small}
where $R$ is the number of modalities; $J_{inter}^{ij}$ represents the inter triplet loss between $i$th modality and $j$th modality, $i\not= j$ represents different modalities;
$J_{intra}^i$ represents the intra triplet loss of $i$th modality; $\mathbf{B}^i$ is the hash codes of $i$th modality; $\mathbf{H}^i\in\mathbb{R}^{k\times N}$
denotes the hash codes of $i$th modality output from the $i$th neural network. It is straightforward to adjust the iterative algorithm presented above to solve
the new optimization problem.
\subsection{Triplet sample}
Due to the limitations in memory size, it is impossible to load all the triplets one time for every instance into the memory to calculate the gradient. A feasible way is
to train the networks iteratively in mini-batch matter, in other words, in each iteration, we select a subset of the triplets to calculate the gradient and then to
update the network parameters. Specifically, we randomly select $P$ anchor instances in each iteration, for each anchor instances, we randomly select $M_1$ positive
instances similar to the anchor instance and $M_2$ negative instances dissimilar to the anchor instance, so we can construct $P\times M_1\times M_2$ triplets to train the networks in each
iteration. And with the increasing number of iterations being executed, the sampled triplets can cover all the possible triplet patterns, ensuring the model to
converge to a Local optimal value.
\section{Experiments and Discussions}
To verify the effectiveness of the proposed TDH method, we conduct a set of the experiments on two widely-used cross-modal datasets. In the following part, we
first introduce our two benchmark datasets and evaluation protocol, then discuss the parameter settings. Finally, we compare our proposed method with several
state-of-the-art hashing methods and analyze the results.

\subsection{Datasets}
\textbf{MIRFlickr25k}~\cite{a15}: This dataset consists of 25015 images collected from the Flickr website. Each image is associated with several textual tags.
Hence, each instance is an image-text pair and annotated with at least one of 24 unique labels. In our experiment, we retain only those instances which have at
least 20 textual tags and remove the others, leaving 20015 instances available. For each instance, the textual instance is represented as a 1386-dimensional
bag-of-words vector. For the shallow methods, the hand-crafted feature for each image is represented as a 512-dimensional SIFT feature vector. For deep hashing
methods, the raw pixels can be directly used as the image modality inputs. Instances are considered to be similar if they share at least one label; otherwise,
they are considered to be dissimilar.
\par\textbf{NUS-WIDE}~\cite{a17}: This dataset contains 269648 web images with 81 concept labels, also collected from Flickr. Each image is associated with some
textual tags. This is a multi-label dataset where each instance is tagged with one or more concepts. However, only a few instances exist for some concepts.
Consequently, we select the 10 most frequently occurring concepts and are thus left with 186577 image-text pairs for our experiment. For each instance, the textual
instance is represented as a 1000-dimensional bag-of-words vector. For the shallow methods, the hand-crafted feature for each image is represented as a
500-dimensional bag-of-visual-words SIFT feature vector. We use the raw pixels directly as the image modality input for deep hashing methods. Instances are
considered to be similar if they share at least one label; otherwise, they are considered to be dissimilar.
\begin{table}
\caption{Comparison to baselines with CNN-F features on MIRFLICKR-25K in terms of MAP. Best accuracy is shown in boldface.}
\label{tab:flickr_CNNF}
\center
\tabcolsep8pt
\arrayrulewidth1pt
\begin{tabular}{|c|c|c|c|c|}
\hline
\multirow{2}{*}{Task/MIRFlickr25k}&
\multirow{2}{*}{Methods}&
\multicolumn{3}{c|}{Code Length}\\
\cline{3-5}
&&16 bits&32 bits&64 bits\\
\hline
\hline
\multirow{10}{*}{\tabincell{c}{Image Query\\v.s.\\Text Database}}&CMFH~\cite{a2}&0.5451&0.5455&0.5451\\
\cline{2-5}
&SCM~\cite{a6}&0.6095&0.6139&0.6143\\
\cline{2-5}
&LSSH~\cite{a3}&0.5712&0.5822&0.5880\\
\cline{2-5}
&STMH~\cite{a20}&0.5944&0.5948&0.6047\\
\cline{2-5}
&CVH~\cite{a5}&0.5378&0.5378&0.5378\\
\cline{2-5}
&SePH~\cite{a16}&0.6984&0.7048&0.7086\\
\cline{2-5}
&DCMH~\cite{a7}&0.7056&0.7035&0.7140\\
\cline{2-5}
&PRDH~\cite{a11}&0.6819&0.6917&0.6913\\
\cline{2-5}
&\textbf{TDH}&\textbf{0.7110}&\textbf{0.7228}&\textbf{0.7289}\\
\hline
\hline
\multirow{10}{*}{\tabincell{c}{Text Query\\v.s.\\Image Database}}&CMFH~\cite{a2}&0.5354&0.5353&0.5352\\
\cline{2-5}
&SCM~\cite{a6}&0.6316&0.6349&0.6360\\
\cline{2-5}
&LSSH~\cite{a3}&0.5687&0.5707&0.5689\\
\cline{2-5}
&STMH~\cite{a20}&0.5915&0.5931&0.6084\\
\cline{2-5}
&CVH~\cite{a5}&0.5399&0.5352&0.5412\\
\cline{2-5}
&SePH~\cite{a16}&0.6438&0.6460&0.6518\\
\cline{2-5}
&DCMH~\cite{a7}&0.7311&0.7487&0.7499\\
\cline{2-5}
&PRDH~\cite{a11}&0.7340&0.7397&0.7418\\
\cline{2-5}
&\textbf{TDH}&\textbf{0.7422}&\textbf{0.7500}&\textbf{0.7548}\\
\hline
\end{tabular}
\end{table}
\begin{table}
\caption{Comparison to baselines with CNN-F features on NUS-WIDE in terms of MAP. Best accuracy is shown in boldface.}
\label{tab:NUS-WIDE_CNNF}
\center
\tabcolsep8pt
\arrayrulewidth1pt
\begin{tabular}{|c|c|c|c|c|}
\hline
\multirow{2}{*}{Task/NUS-WIDE}&
\multirow{2}{*}{Methods}&
\multicolumn{3}{c|}{Code Length}\\
\cline{3-5}
&&16bits&32bits&64bits\\
\hline
\hline
\multirow{10}{*}{\tabincell{c}{Image Query\\v.s.\\Text Database}}&CMFH~\cite{a2}&0.3552&0.3549&0.3545\\
\cline{2-5}
&SCM~\cite{a6}&0.4561&0.4664&0.4697\\
\cline{2-5}
&LSSH~\cite{a3}&0.4425&0.4457&0.4539\\
\cline{2-5}
&STMH~\cite{a20}&0.5269&0.5210&0.5461\\
\cline{2-5}
&CVH~\cite{a5}&0.3671&0.3671&0.3672\\
\cline{2-5}
&SePH~\cite{a16}&0.6224&0.6469&0.6609\\
\cline{2-5}
&DCMH~\cite{a7}&0.6141&0.6167&0.6427\\
\cline{2-5}
&PRDH~\cite{a11}&0.5874&0.6154&0.6232\\
\cline{2-5}
&\textbf{TDH}&\textbf{0.6393}&\textbf{0.6626}&\textbf{0.6754}\\
\hline
\hline
\multirow{10}{*}{\tabincell{c}{Text Query\\v.s.\\Image Database}}&CMFH~\cite{a2}&0.3724&0.3723&0.3722\\
\cline{2-5}
&SCM~\cite{a6}&0.4561&0.4707&0.4799\\
\cline{2-5}
&LSSH~\cite{a3}&0.4153&0.4295&0.4415\\
\cline{2-5}
&STMH~\cite{a20}&0.5089&0.5160&0.5420\\
\cline{2-5}
&CVH~\cite{a5}&0.3642&0.3596&0.3568\\
\cline{2-5}
&SePH~\cite{a16}&0.5658&0.5596&0.6016\\
\cline{2-5}
&DCMH~\cite{a7}&0.6591&0.6487&\textbf{0.6847}\\
\cline{2-5}
&PRDH~\cite{a11}&0.6303&0.6432&0.6568\\
\cline{2-5}
&\textbf{TDH}&\textbf{0.6647}&\textbf{0.6758}&0.6803\\
\hline
\end{tabular}
\end{table}
\subsection{Evaluation Protocol}
For a cross-modal hashing retrieval, we generally conduct studies for two typical tasks: text query in image dataset and image query in textual dataset.
To evaluate the performance of the proposed TDH method, we adopt three widely used evaluation criteria: the precision-recall curve, the top\emph{N}-precision
curve and Mean Average Precision (MAP)~\cite{a18}. The first criterion is based on hash look up, which constructs a look up table using the hash codes and
returns all instances within a given Hamming radius of the query instance. The last two criteria belong to Hamming ranking, which ranks the instances in the
retrieval set according to the Hamming distance between the given query instance and the returned nearest neighbors from the top of the ranking list. These
three criteria may be explained in more detail as follows:
\par MAP is one of the most widely-used measures for cross-modal hashing. To calculate the MAP value, we first evaluate the Average Precision (AP). Given a
query instance, we obtain a ranking list of $R$ retrieved results, and the value of its AP is defined as:
\begin{equation}
AP = \frac{1}{N}\sum\nolimits^{R}\limits_{r=1}p(r)\delta(r),
\end{equation}
where $N$ is the number of relevant instances in the retrieved set, $p(r)$ denotes as the precision of the top $r$ retrieved instances, and $\delta(r)=1$ if
the $r$th retrieved result is relevant to the query instances, otherwise, $\delta(r)=0$. We then obtain the MAP measure by averaging the APs of all the queries
in the query set. The larger the MAP, the better the retrieval performance. The top\emph{N}-precision curve reflects the changes in precision according to the
number of retrieved instances. Finally, the precision-recall curve reflects the precision at different recall levels and can be obtained by varying the Hamming
radius of the retrieved instances in a certain range in order to evaluate the precision and recall.
\begin{table*}
\caption{Comparison of different loss functions in terms of MAP. Best accuracy is shown in boldface.The code length is 16.}
\label{tab:Loss}
\center
\tabcolsep8pt
\arrayrulewidth1pt
\begin{tabular}{|c|c|c|c|c|c|}
\hline
\multicolumn{2}{|c|}{Dataset/Loss}&$J_{intra}+J_{inter}$&$J_{inter}+J_{re}$&$J_{intra}+J_{re}$&$J_{intra}+J_{inter}+J_{re}$\\
\hline
\multirow{2}{*}{MIRFlickr25k}&$I\rightarrow T$&0.7104&0.6670&0.5800&\textbf{0.7110}\\
\cline{2-6}
&$T\rightarrow I$&0.7414&0.6830&0.5938&\textbf{0.7422}\\
\hline
\multirow{2}{*}{NUSWIDE}&$I\rightarrow T$&0.6245&0.5787&0.3750&\textbf{0.6393}\\
\cline{2-6}
&$T\rightarrow I$&0.6597&0.6050&0.4058&\textbf{0.6647}\\
\hline
\end{tabular}
\end{table*}
\subsection{Baselines}
We compare our proposed TDH method with eight state-of-the-art cross-modal hashing methods, namely CMFH~\cite{a2}, SCM~\cite{a6}, LSSH~\cite{a3}, STMH~\cite{a20},
CVH~\cite{a5}, SePH~\cite{a16}, DCMH~\cite{a7} and PRDH~\cite{a11}, where DCMH and PRDH are deep methods and the rest are shallow methods. And the deep networks of
DCMH and PRDH are same as our proposed method. They also utilize the CNN-F network as the deep architecture for the image modality, and also utilize Multi-Layers
Perceptron (MLP) as the deep architecture for the textual modality. Source code for most
baselines was kindly provided by the corresponding authors, except for CMFH. SePH is a kernel-based method, for which we use RBF kernel and select 500 points as
kernel bases in accordance with the authors' suggestions. All other parameters for all baselines are set according to the suggestion provided in the original
papers.

\subsection{Settings}
For the MIRFlickr25k dataset, we select 2000 instances as the query set and the rest of the instances as the retrieval set. To reduce computational cost, the
training set consists of 5000 instances which are sampled randomly from the retrieval set. For the NUS-WIDE dataset, we take 1866 instances as the query set and
the remaining instances as the retrieval set. We also randomly sample 5000 instances from the retrieval set to construct the training set.
\par In our methods, positive instances are those that share at least one common label, while negative instances share no common labels. Triplet labels are also
sampled in every iteration. In the training procedure, triplet labels are used alternatively for image-to-text and text-to-image.
\par For our proposed method, we set the hyper-parameter $\alpha$ according to~\cite{a14}. $\alpha$ is set to half of the hash code length, e.g., 8 for 16-bit hash
codes. We also use a validation set to choose the tradeoff parameters $\gamma$, $\eta$ and $\beta$. According to the results in the validation set, we set
$\gamma=100$, $\eta =50$ and $\beta=1$. In our experiments, we set the batch size of the mini-batch to 128 and the iteration number of the outer-loop in
Algorithm 1 to 500.
\par TDH is implemented using the open source deep learning toolbox MatConvNet~\cite{a23}. All baselines and our method are carried out on a NVIDIA GTX TITAN X GPU
server.
\begin{figure*}
\centering
\includegraphics[width=18cm]{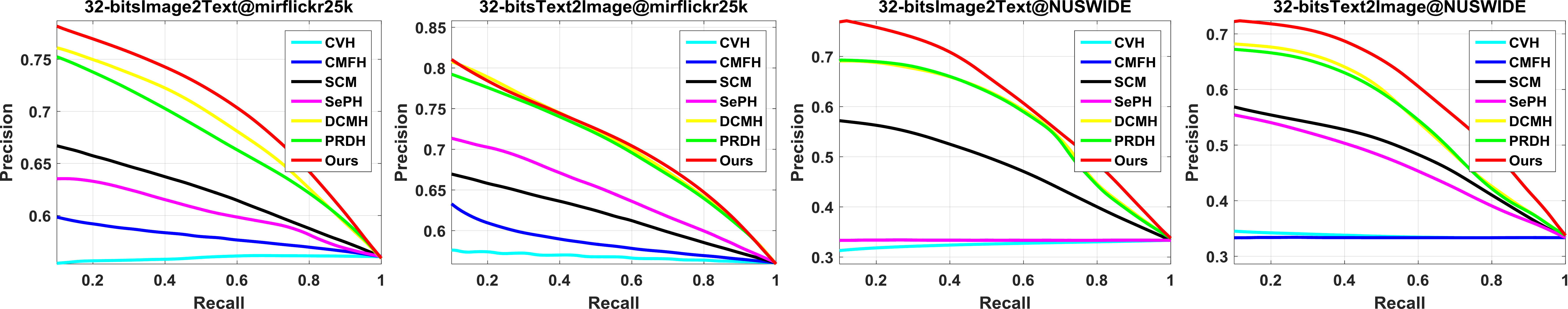}
\caption{Precision-recall curves. The baselines are based on hand-crafted features. The code length is 32.}
\label{fig:32PR_hand}
\end{figure*}
\begin{figure*}
\centering
\includegraphics[width=18cm]{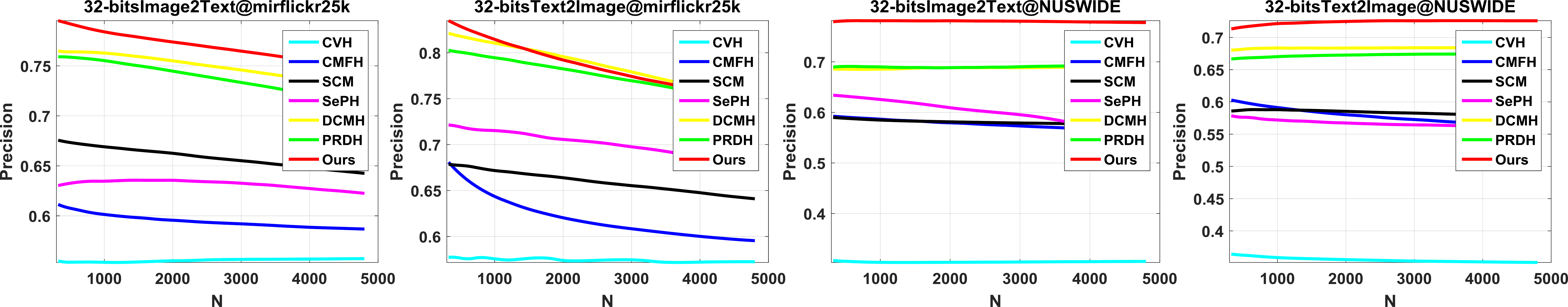}
\caption{Top\emph{N}-precision curves. The baselines are based on hand-crafted features. The code length is 32.}
\label{fig:32topN_hand}
\end{figure*}
\begin{figure*}
\centering
\includegraphics[width=18cm]{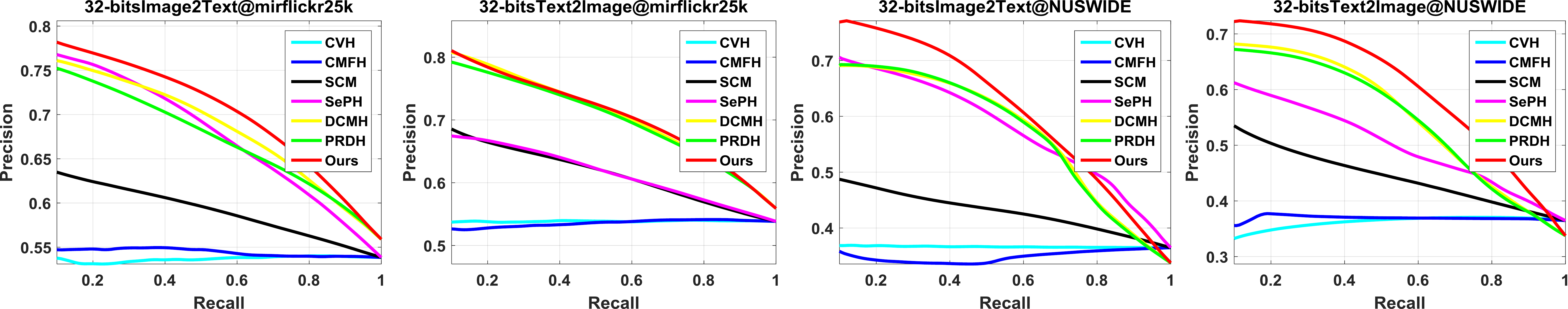}
\caption{Precision-recall curves. The baselines are based on CNN-F features. The code length is 32.}
\label{fig:32PR_CNNF}
\end{figure*}
\begin{figure*}
\centering
\includegraphics[width=18cm]{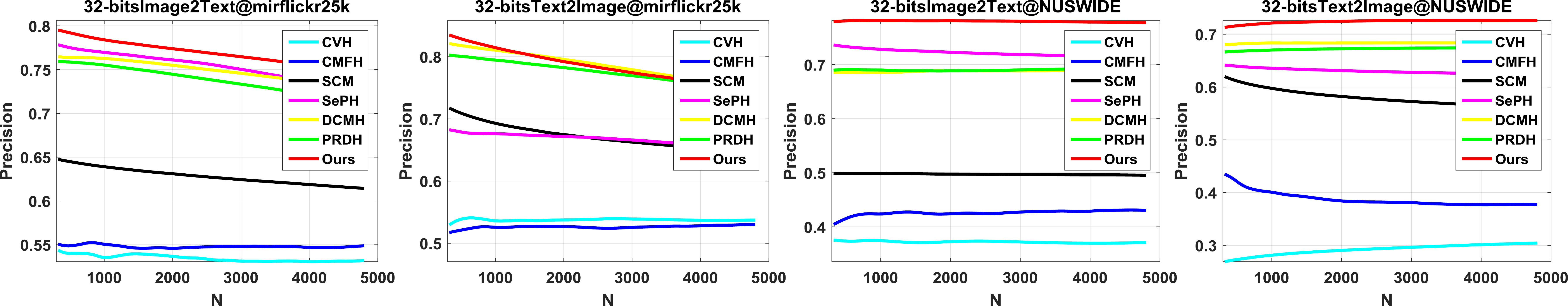}
\caption{Top\emph{N}-precision curves. The baselines are based on CNN-F features. The code length is 32.}
\label{fig:32topN_CNNF}
\end{figure*}
\begin{figure*}
\centering
\includegraphics[width=12cm]{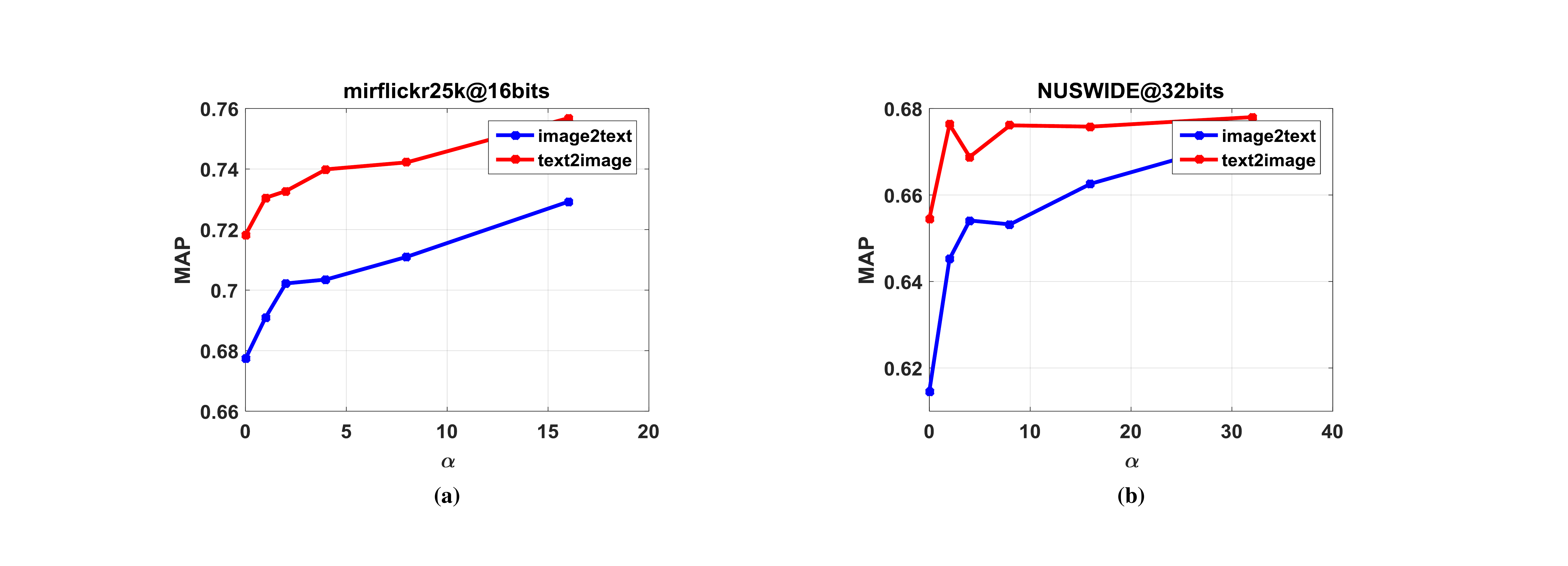}
\caption{The impact of the hyper-parameter $\alpha$. (a) The MAP on the MIRFlickr25k dataset with 16 bits under different values of $\alpha$. (b) The MAP on the NUS-WIDE dataset with 32 bits under different values of $\alpha$.}
\label{fig:alpha}
\end{figure*}
\subsection{Performance Comparisons and Disccussions}
Table~\ref{tab:flickr_hand} and Table~\ref{tab:NUS-WIDE_hand} present, the MAP values of TDH, deep methods and other shallow methods with hand-crafted features for
text query image task and image query text task for the MIRFlickr25k and NUS-WIDE datasets respectively. We set the length of hash codes to be 16 bits, 32 bits
and 64 bits. From the results, we find that the proposed TDH method is superior to all the baselines on different retrieval tasks. To further prove the
effectiveness of our proposed TDH method on cross-modal retrieval, we exploit the CNN-F deep network~\cite{a10} which is the same as the initial CNN of the image
modality in TDH to extract CNN-F features. All shallow methods are trained based on these CNN-F features. Table~\ref{tab:flickr_CNNF} and
Table~\ref{tab:NUS-WIDE_CNNF} report the MAP values of TDH and all baselines with CNN-F features  for text query image task and image query text task for the
MIRFlickr25k and NUS-WIDE datasets respectively. From the results, we can find that the proposed TDH method is superior to all the baselines with CNN-F features on
different retrieval tasks. Compared with the pairwise labels, triplet labels significantly boost the correlation among the three instances allowing to more semantic information
to be explored. We add the graph regularization in Hamming space into the model to preserve the similarities among cross-modal original features. We also find that
text query task always has better results than image query task, which shows that it is more difficult to find the semantic similarities between texts and images
via image query. This may mean that the hidden semantic information in an image is more difficult to capture than that in text. From the MAP results, we find that
the increasing in MAP for the text query image task is less than that for image query text task, and that even the MAP of TDH is inferior to that of DCMH, with 64
bits on the NUS-WIDE dataset. Under the triplet loss, the image network reaches convergence more quickly than the textual network. Accordingly, at the same number
of iteration,  the textual network may learn more information.
\par Fig.~\ref{fig:32PR_hand} and Fig.~\ref{fig:32topN_hand} show the precision-recall curves and top\emph{N}-precision curves on two datasets with 32-bit hash
codes, where the shallow methods use hand-crafted features. The curves of TDH are always higher than all other methods as can be seen from the change in the
abscissa. These results are consistent with the MAP evaluation. In short, the proposed TDH method outperforms all of the baselines. Fig.~\ref{fig:32PR_CNNF} and
Fig.~\ref{fig:32topN_CNNF} show the precision-recall curves on two datasets with 32-bit hash codes, where the shallow methods use CNN-F features. We can also find
that TDH can outperform all the other baselines with CNN-F features.
\par Table~\ref{tab:Loss} reports the MAP values of TDH on two datasets with 16-bit hash codes, where different loss functions contain different loss
components, and reflect their individual effect in the objective function. It can be seen that $J_{inter}$ has the biggest effect, followed by $J_{intra}$ and $J_{re}$
has least effect in the objective function. The inter-modal triplet embedding loss is used to capture the heterogeneous correlation across different modalities, the
intra-modal triplet loss explores the discriminative abilities of hash codes, and the regularization loss can enforce the adjacency consistency to guarantee that
the learned hash codes can preserve the original cross-modal semantic similarities in Hamming space. So for the objective function, these three components are
indispensable.
\par We also implement a parameter sensitivity experiment for the hyper-parameter $\alpha$. Fig.~\ref{fig:alpha} shows the impact of $\alpha$ at 16 bits on the
MIRFlickr25 dataset and 32 bits on the NUS-WIDE dataset. We can see that the larger the value of $\alpha$, the better the performance. Finally, $\alpha$  is set
to half of the hash code length.
\section{Conclusion}
In this paper, we propose a novel hashing method dubbed triplet-based deep hashing (TDH) network for cross-modal retrieval. The proposed TDH method learns an
end-to-end framework to integrate feature learning and hash code learning. Triplet labels are exploited as supervised information to capture relative semantic
correlation between heterogeneous data from different modalities. In addition, we establish the loss function from the inter-modal view and the intra-modal view
to enhance the discriminative ability of hash codes. Finally, We introduce graph regularization into Hamming space to preserve the original semantic similarities
of the learned hash codes. Experimental results on two popular datasets have demonstrated that our TDH method outperforms several state-of-the-art approaches.
\section*{Acknowledgment}
This work was supported in part by the National Natural Science Foundation of China(61572388, 61703327), the Key R\&D Program-The Key Industry
Innovation Chain of Shaanxi(2017ZDCXL-GY-05-04-02, 2017ZDCXL-GY-05-04-02) and Australian Research Council Projects FL-170100117, DP-180103424, LP-150100671.

%


\ifCLASSOPTIONcaptionsoff
  \newpage
\fi

\bibliographystyle{ieeetr}
\bibliography{bibfile}
\vspace{-0.5cm}
\begin{IEEEbiography}[{\includegraphics[width=1in,height=1.25in,clip,keepaspectratio]{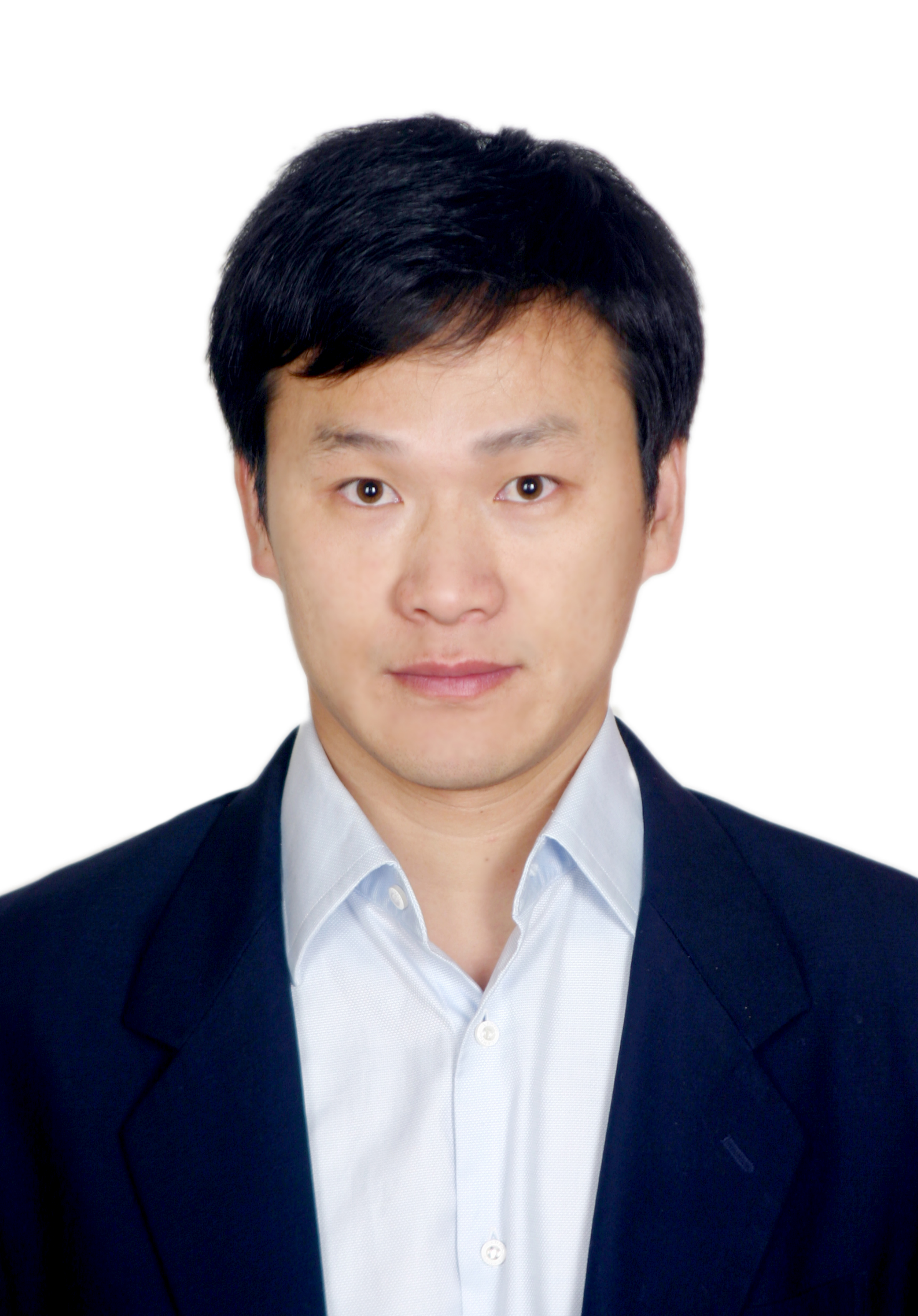}}]{Cheng Deng}
(S'09) received the B.E., M.S., and Ph.D. degrees in Signal and Information Processing from Xidian University, Xi¡¯an, China. He is currently a Full Professor with the School of Electronic Engineering at Xidian
University. His research interests include computer vision, multimedia processing and analysis, and information hiding. He is the author and coauthor of more than 50 scientific articles at top venues,
including IEEE TNNLS, TMM, TCYB, TSMC, TIP, ICCV, CVPR, IJCAI, and AAAI.
\end{IEEEbiography}
\vspace{-0.5cm}
\begin{IEEEbiography}[{\includegraphics[width=1in,height=1.25in,clip,keepaspectratio]{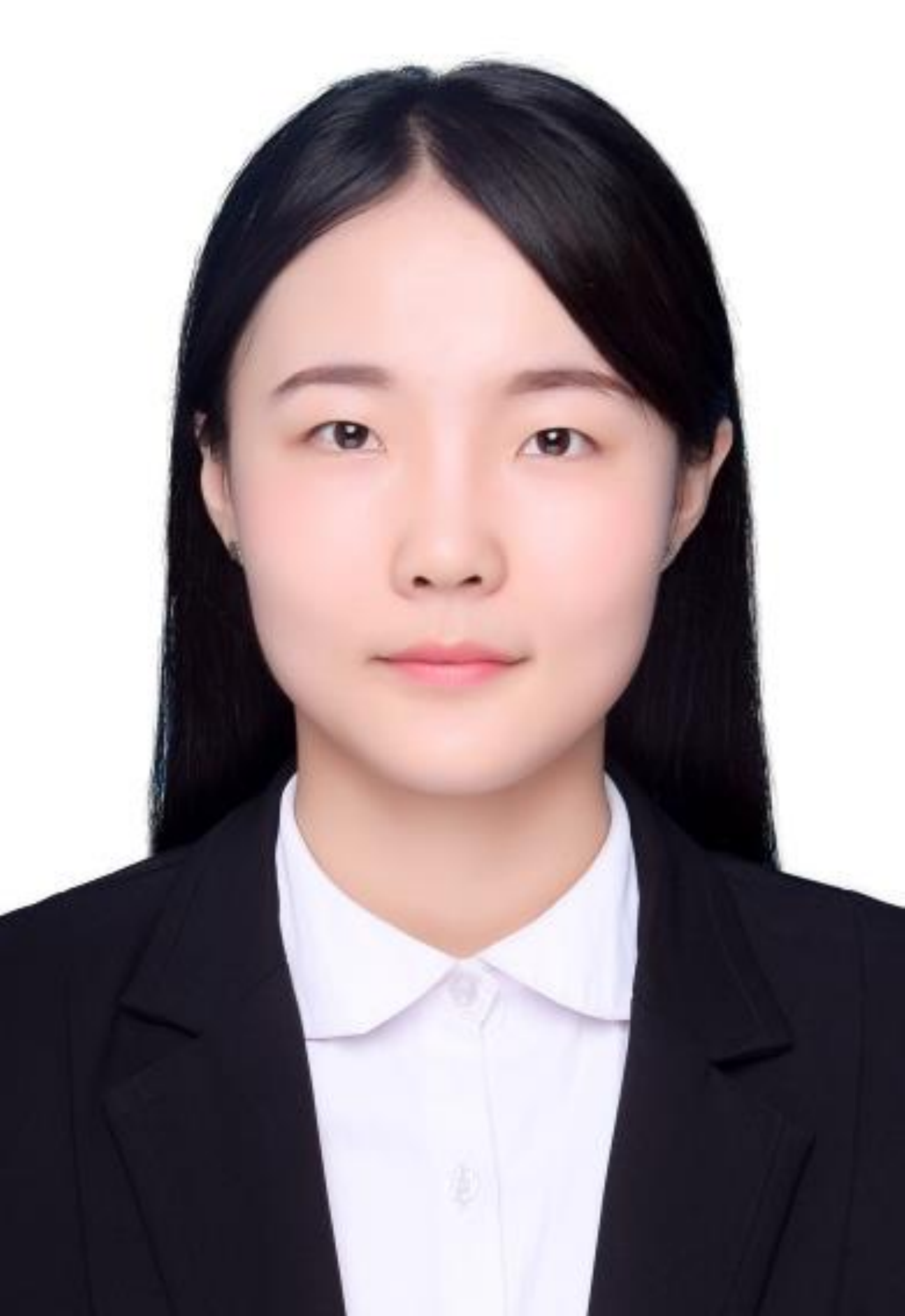}}]{Zhaojia Chen}
received the B.Sc degree from Xidian
University, Xi¡¯an, China, in 2015, and is currently
working toward the M.S. degree in electronic engineering at Xidian University.
Her research interests focus on large-scale multimedia retrieval.
\end{IEEEbiography}
\vspace{-0.5cm}
\begin{IEEEbiography}[{\includegraphics[width=1in,height=1.25in,clip,keepaspectratio]{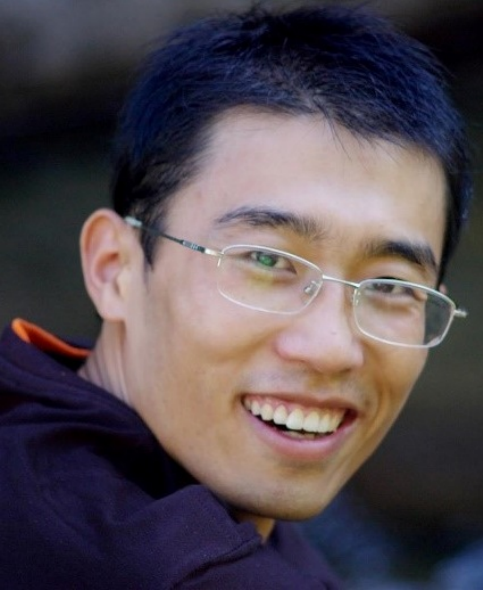}}]{Xianglong Liu}
received the BS and Ph.D degrees in computer science from Beihang University, Beijing, in 2008 and 2014. From 2011 to 2012, he visited the Digital Video and Multimedia (DVMM) Lab, Columbia University as a joint
Ph.D student. He is now an assistant professor with Beihang University. His research interests include machine learning, computer vision and multimedia information retrieval.
\end{IEEEbiography}
\vspace{-0.5cm}
\begin{IEEEbiography}[{\includegraphics[width=1in,height=1.25in,clip,keepaspectratio]{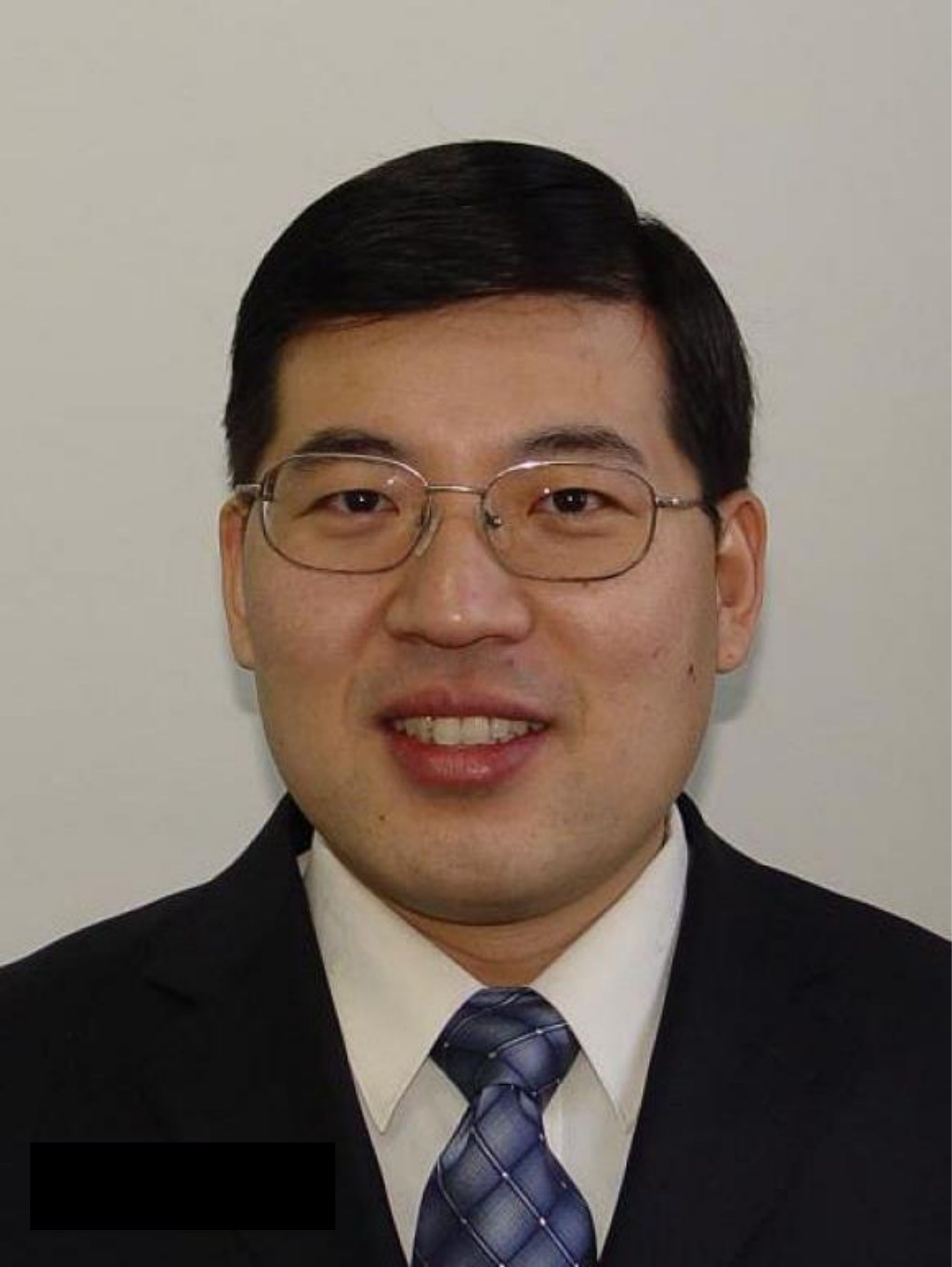}}]{Xinbo Gao}
(M'02-SM'07) received the B.Eng., M.Sc., and Ph.D. degrees from Xidian University, Xi¡¯an, China, in 1994, 1997, and 1999, respectively, all in signal and information processing. From 1997
to 1998, he was a Research Fellow with the Department of Computer Science, Shizuoka University, Shizuoka, Japan. From 2000 to 2001, he was a PostDoctoral Research Fellow with the Department of
Information Engineering, The Chinese University of Hong Kong, Hong Kong. Since 2001, he has been
with the School of Electronic Engineering, Xidian University. He is currently a Cheung Kong Professor of the Ministry of Education, a Professor of Pattern Recognition and Intelligent System, and the
Director of the State Key Laboratory of Integrated Services Networks, Xi¡¯an. His current research interests include multimedia analysis, computer vision,
pattern recognition, machine learning, and wireless communications. He has published six books and around 200 technical articles in refereed journals
and proceedings. He is on the Editorial Boards of several journals, including Signal Processing (Elsevier) and Neurocomputing (Elsevier). He served as the
General Chair/Co-Chair, Program Committee Chair/Co-Chair, or PC Member for around 30 major international conferences. He is a fellow of the Institute
of Engineering and Technology and the Chinese Institute of Electronics.
\end{IEEEbiography}
\begin{IEEEbiography}[{\includegraphics[width=1in,height=1.25in,clip,keepaspectratio]{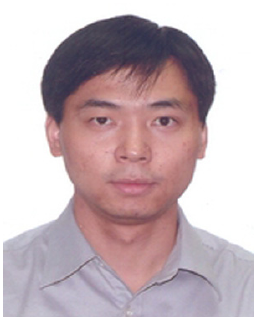}}]{Dacheng Tao}
(F'15) is Professor of Computer Science and ARC Laureate Fellow in the School of Information Technologies and the Faculty of Engineering and Information Technologies,
and the Inaugural Director of the UBTECH Sydney Artificial Intelligence Centre, at the University of Sydney. He mainly applies statistics and mathematics to Artificial
Intelligence and Data Science. His research interests spread across computer vision, data science, image processing, machine learning, and video surveillance. His research
results have expounded in one monograph and 500+ publications at prestigious journals and prominent conferences, such as IEEE T-PAMI, T-NNLS, T-IP, JMLR, IJCV, NIPS, ICML,
CVPR, ICCV, ECCV, ICDM; and ACM SIGKDD, with several best paper awards, such as the best theory/algorithm paper runner up award in IEEE ICDM¡¯07, the best student paper award
in IEEE ICDM¡¯13, the distinguished student paper award in the 2017 IJCAI, the 2014 ICDM 10-year highest-impact paper award, and the 2017 IEEE Signal Processing Society Best
Paper Award. He received the 2015 Australian Scopus-Eureka Prize, the 2015 ACS Gold Disruptor Award and the 2015 UTS Vice-Chancellor¡¯s Medal for Exceptional Research.
He is a Fellow of the IEEE, AAAS, OSA, IAPR and SPIE.
\end{IEEEbiography}
\end{document}